# Investigating student understanding of quantum entanglement

Antje Kohnle and Erica Deffebach

*University of St Andrews, School of Physics and Astronomy,*
*North Haugh, St Andrews, KY16 9SS, United Kingdom*

**Abstract.** Quantum entanglement is a central concept of quantum theory for multiple particles. Entanglement played an important role in the development of the foundations of the theory and makes possible modern applications in quantum information technology. As part of the QuVis Quantum Mechanics Visualization Project, we developed an interactive simulation *Entanglement: The nature of quantum correlations* using two-particle entangled spin states. We investigated student understanding of entanglement at the introductory and advanced undergraduate levels by collecting student activity and post-test responses using two versions of the simulation and carrying out a small number of student interviews. Common incorrect ideas found include statements that all entangled states must be maximally entangled (i.e. show perfect correlations or anticorrelations along all common measurement axes), that the spins of particles in a product state must have definite values (cannot be in a superposition state with respect to spin) and difficulty factorizing product states. Outcomes from this work will inform further development of the QuVis *Entanglement* simulation.

**PACS:** 01.50.ht, 03.65.Ud

## I. INTRODUCTION

For classical composite systems each of the subsystems has well-defined properties. For quantum-mechanical composite systems, there exist states for which the wave function of the composite system is known, but the subsystems cannot be described in terms of individual wave functions and thus cannot be described separately. Such states for which the total wave function is not the product of individual wave functions, e.g. is not factorizable, are called entangled. Thus, entangled states are not product states.

Schrödinger famously stated that "entanglement is not one but rather the characteristic trait of quantum mechanics" [1]. A remarkable feature is that two entangled quantum particles can show correlations in measurement outcomes that are not reproducible by classical models. Through this feature, entanglement has important physical consequences including the Bell inequalities and applications in teleportation, quantum computing and cryptography [2-4].

Given the key role of entanglement in the description of quantum systems of multiple particles, helping students come to a correct understanding of entanglement is an important instructional goal. Existing studies of student difficulties in quantum mechanics cover various topics but do not include entanglement [5]. As part of the QuVis Quantum Mechanics Visualization Project [6], we have developed an interactive simulation *Entanglement: The nature of quantum correlations* (henceforth referred to as the *Entanglement s*imulation) using two-particle entangled spin states. The simulation allows students to explore experimental outcomes for various input states and easily switch between product states and entangled states [7]. In this study, we investigated student understanding of entanglement using two versions of the simulation. Our aims in this work are to assess what common incorrect ideas persist after instruction and *Entanglement* simulation use. Outcomes will inform further development of this simulation.

## II. METHODOLOGY

The QuVis *Entanglement* simulation does not require the mathematical formalism of tensor products and is aimed at the introductory and advanced undergraduate levels. A screenshot of the revised, second version of the simulation is shown in Fig. 1. The simulation shows a source of particle pairs in the middle of two Stern-Gerlach apparatuses (SGAs), which can be jointly rotated along two orthogonal axes, denoted X and Z. The states $|\uparrow_A\rangle$ and $|\downarrow_B\rangle$ refer to spin-up and spin-down states along the Z-axis for particles A and B respectively. Students can choose between different input states (left panel in Fig. 1) and send particle pairs through the experiment. The individual and paired measurement outcomes and the correlation coefficient are shown (middle and right panels in Fig. 1). The correlation coefficient is the average value of the product of the two measurement outcomes, defined as +1 when the deflections are the same and −1 when the deflections are opposite. A correlation coefficient of +1 implies perfect correlation, of −1 perfect anticorrelation. Besides the "Controls" view shown in Fig. 1, the simulation also includes explanatory texts in the "Introduction" and "Step-by-step Explanation" views.

In the initial first version of the simulation, users could only choose between three fixed input states, including one entangled state (see Fig. 2). Users could choose to display the states in the X or Z basis, and show the states as products of the two individual particle states or in expanded form as depicted in Fig. 2. Due to difficulties found (see section Outcomes), the revised simulation shows the first two states in both the X and Z bases. It allows users to create their own state by putting together different two-particle spin states, as shown in the lower-left panel of Fig. 1. The revised version also allows users to choose between two different notations for the spin states.

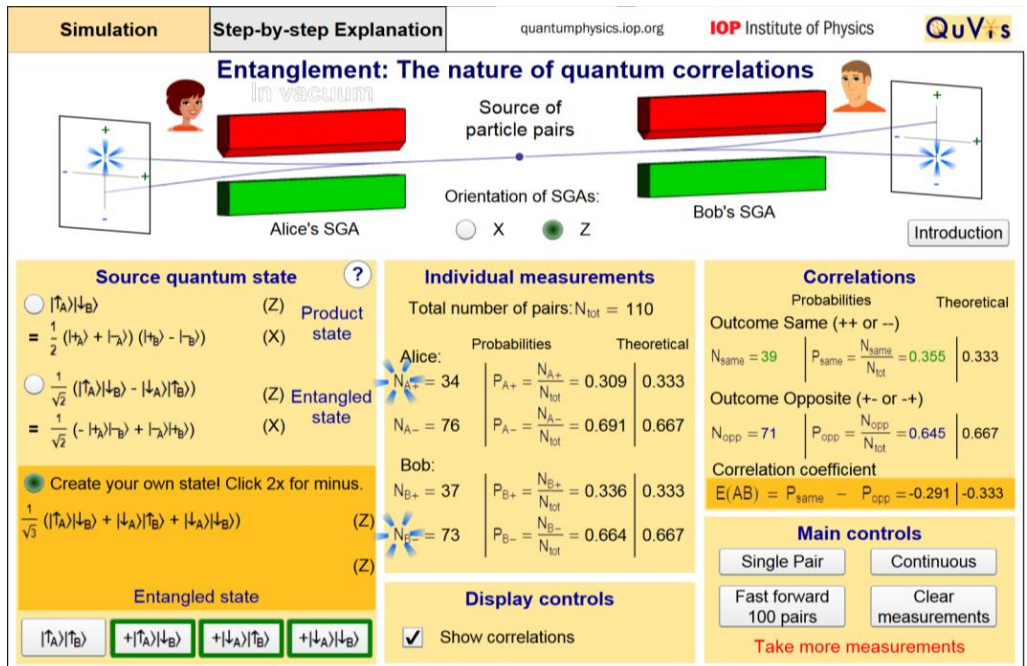

**FIG 1.** A screenshot of the "Controls" view of the revised version of the *Entanglement* simulation.

The accompanying activity to the revised simulation shown in Fig. 1 asks students to explain the observed individual and paired measurement outcomes and the correlation coefficient for the first input state (a product state) considering both orientations of the Stern-Gerlach apparatuses. Students are asked to rewrite this state in the X basis and explain why this state is a product state. The activity then asks students to choose the second input state (a maximally entangled state that always has opposite outcomes), and to compare and contrast the previous product state and this entangled state in terms of measurement outcomes. Students are then asked to use the "Create your own state" option to create entangled states with different correlations, including an entangled state for which there are no correlations in the X and Z bases. Students are also asked whether a product state implies that the spins of particles have definite values. The activity to the original version of the simulation was similar, but did not include the parts where students create their own states as this option was not available (see Fig. 2).

We collected written responses to the *Entanglement* simulation activity and in cases also written post-test responses using the original and the revised versions of the simulation (see Table 1). The 2015 post-test questions are shown in Fig. 3. For post-test question 1, states a) and d) are entangled. For post-test question 2, only statement II is correct. The post-test questions are multiple-choice, but students were asked to explain their reasoning for each question. Trials using the simulation were carried out in an introductory quantum physics course (often the first university course in quantum physics that students take, similar to a US Modern Physics course) and a senior-level Advanced Quantum Mechanics course, both at the University of St Andrews.

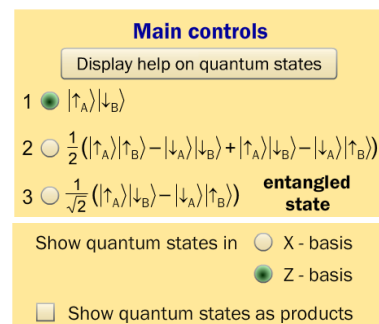

**FIG 2.** A screenshot showing parts of the original version of the *Entanglement* simulation. Only three fixed input states are available. States are shown in either the X or Z basis, not both simultaneously.

Revisions were incorporated into the simulation prior to the 2015 trial based on analysis of the 2013 and 2014 trials. For the advanced course, post-tests were given in the lecture directly after homework submission. For the introductory course, the post-test was completed in the last minutes of a 50-minute computer workshop. We also carried out interviews in 2015 with five students from the introductory level a few days after the *Entanglement* simulation was used. These interviews confirmed our interpretation of written student reasoning.

**TABLE 1.** The table shows the number N of activity and post-test responses where applicable collected from courses at the University of St Andrews. The 2013 and 2014 courses used the initial version of the simulation, the 2015 course used the revised version.

| Level | Year | N | Post-test | Simulation use |
|---|---|---|---|---|
| Introductory | 2013 | 59 | none | Computer workshop |
| Advanced | 2014 | 24 | Post-test | Homework |
| Introductory | 2015 | 79 | Post-test | Computer workshop |

---
1) Consider the following two-particle states of a pair of spin ½ particles. The notations $|\uparrow\rangle$ and $|\downarrow\rangle$ refer to spin states with z-component of spin $S_z = +\hbar/2$ and $S_z = -\hbar/2$ respectively. The numerical factors are needed for normalization.
Which of these states is/are entangled? Choose one or more.

a) $1/\sqrt{2} \, (|\uparrow_A\rangle|\uparrow_B\rangle + |\downarrow_A\rangle|\downarrow_B\rangle)$
b) $1/2 \, (|\uparrow_A\rangle|\uparrow_B\rangle + |\downarrow_A\rangle|\downarrow_B\rangle + |\uparrow_A\rangle|\downarrow_B\rangle + |\downarrow_A\rangle|\uparrow_B\rangle)$
c) $1/\sqrt{2} \, (|\downarrow_A\rangle|\uparrow_B\rangle - |\downarrow_A\rangle|\downarrow_B\rangle)$
d) $1/2 \, (|\uparrow_A\rangle|\uparrow_B\rangle + |\downarrow_A\rangle|\downarrow_B\rangle - |\uparrow_A\rangle|\downarrow_B\rangle + |\downarrow_A\rangle|\uparrow_B\rangle)$

2) Do you agree or disagree with the following statements:

I. If the source is emitting entangled particle pairs, then there will be perfect correlation along both X and Z, or perfect anticorrelation along both X and Z.

II. If Alice and Bob find perfect anticorrelation along both X and Z, they know that the source must be emitting entangled particle pairs.

---

**FIG 3.** The 2015 post-test questions. Question 1 of the 2014 post-test only included options a), b) and c). Question 2 was only used in 2015.

We marked written responses to the activity questions as correct, partially correct, incorrect and unanswered and compiled the fractions of each per question. For the post-tests, we analyzed students' choices and reasoning in assessing correctness of responses and incorrect ideas, with both reasoning and choices needing to be correct for a response to be coded as correct. We coded incorrect and partially correct responses using an emergent coding scheme, using the same codes for the activity responses and the post-test responses. The 2013 and 2014 activity responses and post-test responses including reasoning were coded by both authors and checked for inter-rater reliability. Categories with disagreement were discussed and revised until high inter-rater reliability was achieved (88% agreement for the 2013 data and 86% for the 2014 data). Due to time constraints, the 2015 data was only coded by one author and checked for consistency by the other author using a subset of the data.

In the lectures, the introductory course only discussed a maximally-entangled two-particle state, and did not define entangled states in terms of not factorizable states. Thus, introductory students were learning about product states and non-maximally entangled states from the simulation alone. In the advanced course lectures, entanglement was introduced via states that are not product states but the focus was primarily on maximally-entangled states and the density matrix formalism.

### III. OUTCOMES

In what follows, we discuss common incorrect ideas found in student reasoning. Frequencies across the different levels and years are summarized in Table 2.

**A. For an entangled state, if you know the measurement outcome of one particle, the outcome of the other particle is completely determined. Entangled states show either perfect correlations or perfect anticorrelations along all common axes.** This idea incorrectly assumes that all entangled states are maximally-entangled, i.e. show either perfect correlation or perfect anticorrelation along all common measurement axes. A typical student response describing entanglement illustrating this idea is "*If you make a measurement on one particle, you know the measurement of the other and they have to be either the opposite or either the same.*"

The 2014 post-test question 1 included three states (options a) to c) in Fig. 3, one maximally-entangled state and two product states). Five advanced level students correctly identified the entangled state, but incorrectly reasoned that product states are those for which the outcome of one particle is not fixed when the other is measured. For example, a student reasons "*for a) the two outcomes of the experiment are both A and B measuring the same spin. This means that there is a dependence upon the measurement of A on B and vice versa. Hence a) is an entangled state. For the other states the outcomes can differ in whether A and B measured the same or opposite spin, hence no dependence exists between the measurements. Therefore b) and c) are not entangled.*"

These outcomes led us to develop the "Create your own state" option in the revised simulation (see Fig. 1) used in 2015. The revised activity now asks students to create entangled states that do not exhibit perfect correlations or anticorrelations. However, 28 students (35%) in the 2015 introductory level trial incorrectly agreed with post-test question 2 statement I (see Fig. 3), showing that this incorrect idea persists even after students made use of the revised simulation.

**B. Incorrect properties of product states, e.g. that product states can be entangled states along a different basis, or that product states can also show perfect**

**correlations along all bases.** These ideas are linked with difficulties translating a state from the Z to the X basis. The 2015 activity explicitly asked students to rewrite a state given in the Z basis in the X basis, and asked "If a state is a product state along Z, will it also be a product state along X?" These two questions were amongst the most poorly answered, with 74% and 78% correct respectively. Also, 23 students (29%) in the 2015 post-test incorrectly disagreed with statement II of question 2 (Fig. 3). Of these students, 11 (14%) used reasoning similar to *"[statement] II is not correct because product states can exist where there is perfect anticorrelation or correlation along X and Z."* The 2013 and 2014 data did not include questions testing for this difficulty.

**C. Particles in a product state must have definite spin values (i.e., not be in a superposition of spin states).** For the introductory 2013 course, 6 of 59 students (10%) stated that this is the case in response to a question "Entangled states are not product states. Interpret this statement physically." For example, a student states *"For a product state both particles have a definite value of spin measured along a given axis. For an entangled state both particles do not have well-defined spins although their relative spins are always well-defined"*. In the 2015 activity to the revised simulation, we explicitly asked "Does a product state imply that the spins of the particles have definite values?" 10 of 79 (13%) students incorrectly stated that this is the case. Several answers stated (not seen in the 2013 responses) that at least one of the particles must have a definite spin. For example, a student states *"It implies that at least one half of the particle pair does."* This difficulty was only seen at the introductory level.

**D. Incorrectly stating that a product state is an entangled state, due to difficulties converting a product state written as a sum of two-particle terms into the factorized form as a product of two single-particle states.** In the advanced level course, 5 students (21%) stated on question 1 of the post-test (Fig. 3) that $1/\sqrt{2}\,(|\downarrow_A\rangle|\uparrow_B\rangle - |\downarrow_A\rangle|\downarrow_B\rangle)$ is an entangled state as it could not be factorized, i.e. did not recognize that this is the product state $1/\sqrt{2}\,|\downarrow_A\rangle\,(|\uparrow_B\rangle - |\downarrow_B\rangle)$. In the 2015 post-test 8 students (10%) stated the above state could not be factorized. 13 students (16%) stated that state b) (Fig. 3) is an entangled state as it could not be factorized, whereas this is the product $1/2\,(|\uparrow_A\rangle + |\downarrow_A\rangle)\,(|\uparrow_B\rangle + |\downarrow_B\rangle)$. For the 2015 trial, 27 students in total did not factorize states correctly in the post-test (some responses incorrectly factorized entangled states). The 2013 activity did not include questions assessing this difficulty.

Other difficulties seen with lower frequencies include the incorrect ideas that a quantum state with multiple terms must be an entangled state and that entangled states and mixtures are experimentally indistinguishable. There were also incorrect assignments of correlations to quantum states, e.g. stating that a correlation coefficient of +1 implies the individual outcomes must be completely random.

**TABLE 2.** Frequencies of common difficulties found; codes as in the text. Student numbers are in parentheses.

| Code | Intro 2013 Activity | Advanced 2014 Post-test | Intro 2015 Activity | Intro 2015 Post-test |
|---|---|---|---|---|
| A | 8% (5) | 21% (5) | 33% (26) | 35% (28) |
| B | 0% (0) | 0% (0) | 16% (13) | 14% (11) |
| C | 10% (6) | 0% (0) | 13% (10) | 0% (0) |
| D | 0% (0) | 21 % (5) | 10% (8) | 34% (27) |

## IV. CONCLUSIONS

These findings point to difficulties with the relations between superposition and entanglement (entanglement implies superposition but not vice versa, code C) and perfect correlations / anticorrelations along multiple axes and entanglement (these correlations imply entanglement but not vice versa, codes A and B). Based on these outcomes, we plan to revise the *Entanglement* simulation to include another view where students can change the coefficients in an entangled state to explore the transition between maximal and non-maximal entanglement. We plan to add help texts showing how to convert between a sum of terms and the factorized form for a product state and to translate a state from the Z to the X basis. We plan to add a "Challenges" view with multiple challenges targeting the difficulties found. Future work will aim to elicit underlying reasons for the difficulties found in this study, and evaluate the effectiveness of these further simulation revisions using pre- and post-tests and student interviews.

## ACKNOWLEDGEMENTS

We thank Charles Baily, Christopher Hooley and Natalia Korolkova from the University of St Andrews for incorporating the *Entanglement* simulation into their courses. We gratefully acknowledge all of the students who participated in this study. We thank the UK Institute of Physics for funding the simulation development.